\documentclass[nohyper,12pt,letterpaper]{JHEP3}
\usepackage[dvips]{epsfig}
\usepackage{amsfonts,amssymb}

\newcommand{\be}{\begin{equation}}
\newcommand{\ee}{\end{equation}}
\newcommand{\ben}{\begin{displaymath}}
\newcommand{\een}{\end{displaymath}}
\newcommand{\bea}{\begin{eqnarray}}
\newcommand{\eea}{\end{eqnarray}}
\newcommand{\bean}{\begin{eqnarray*}}
\newcommand{\eean}{\end{eqnarray*}}

\def\l {\lambda}

\newcommand{\adss}[2]{\mbox{$AdS_{#1}\times {S}^{#2}$}}

\newcommand{\commentout}[1]{}

\newcommand{\beq}{\begin{equation}}
\newcommand{\eeq}{\end{equation}}
\newcommand{\beqr}{\begin{displaymath}}
\newcommand{\eeqr}{\end{displaymath}}
\newcommand{\beqa}{\begin{eqnarray}}
\newcommand{\eeqa}{\end{eqnarray}}
\newcommand{\beqar}{\begin{eqnarray*}}
\newcommand{\eeqar}{\end{eqnarray*}}
\newcommand{\cN}{{\cal N}}

\newcommand{\half}{\ensuremath{\frac{1}{2}}}

\newcommand{\bz}{\ensuremath{\bar{z}}}
\newcommand{\bZ}{\ensuremath{\bar{Z}}}

\newcommand{\N}[1]{\ensuremath{\cN=#1}}

\newcommand{\sg}{\ensuremath{\mathrm{sign}}}

\newcommand{\bl}{\ensuremath{\bar{\l}}}

\newcommand{\sech}{\mathop{\mathrm{sech}}\nolimits}

\title{\LARGE Scattering of single spikes}

\author{Riei Ishizeki$^1$, Martin Kruczenski$^1$, Marcus Spradlin$^2$, Anastasia Volovich$^2$  \\
 $^1$Dep. of Physics, Purdue Univ., 525 Northwestern Ave., W. Lafayette, IN 47907 \\
 $^2$Department of Physics, Brown University, Box 1843, Providence, RI 02912 \\
E-mail: \email{rishizek@purdue.edu, markru@purdue.edu, spradlin@het.brown.edu, nastja@het.brown.edu}}

\abstract{We apply the dressing method to a string solution given by a static string wrapped 
around the equator of a three-sphere and find that the result is the single spike solution recently discussed
in the literature. Further application of the method allows the construction of solutions with multiple spikes. 
 In particular we construct the solution describing the scattering of two single spikes and compute the scattering 
phase shift. As a function of the
dressing parameters, the result is exactly the same
as the one for 
the giant magnon, up to non-logarithmic terms. This suggests that the single spikes should be described by an integrable 
spin chain closely related to the one associated to the giant magnons. The field theory interpretation of such spin chain 
however is still unclear. 
}


\begin{document}

\section{Introduction}

 The large-$N$ approximation \cite{largeN} seems to be the most promising approach to gain an analytical understanding of the strong 
coupling regime of gauge theories. This belief is due
in great part to the AdS/CFT correspondence \cite{malda}, which provides a concrete 
example where such an approximation works.  The correspondence argues that at large 't Hooft coupling $\lambda$, four dimensional 
$SU(N)$ \N{4} SYM  theory is described by classical strings in \adss{5}{5}. Quantum effects on the world-sheet are suppressed 
by $\frac{1}{\sqrt{\lambda}}$ and string loop effects by $\frac{1}{N}$. String states can be seen to appear from the field theory \cite{bmn,GKP}
as long gauge invariant operators. 
Since at large $\lambda$ a semi-classical expansion is appropriate, a particularly important role is played by classical 
string solutions \cite{FT,BFST}. In certain limits
these can be directly mapped \cite{kru,KRT,HL} to spin chains which appear also in the 
field theory \cite{MZ} as a way to describe a certain class of long operators. 

For our purpose, some particular solutions recently proposed in \cite{IK,Mo} and called ``single spike solutions'' will be of interest. 
They were found based on previous work \cite{spiky, Ryang, HM} and shown to be closely related in their properties to the giant magnon 
solutions described in \cite{HM} and further analyzed in \cite{Dorey}--\cite{Bozhilov:2007wn}.

 A very useful tool
for understanding the giant magnon solutions is the dressing method\footnote{The dressing method can also 
be applied in the $AdS$ sector, for example it was used in \cite{Jevicki:2007pk} to find new solutions by dressing the solution in \cite{Kruczenski:2002}.}
of \cite{dm1,dm2,dm3} which was shown in \cite{SV,KSV} to provide a simple description of the magnon as well as a method
for constructing
multiple magnon solutions. Superposing magnons is in principle difficult since the problem is non-linear, and becomes possible only due to the integrability 
of the equations of motion. It appears natural to ask if the dressing method can be similarly applied to the study of spike 
solutions and their scattering.

 In this paper we find that, indeed, the dressing method provides for a simple understanding of the single spike solutions and furthermore
allows the construction of new solutions with multiple spikes. Of particular interest are solutions describing the scattering of two 
single spikes from which we can compute the scattering phase shift. This calculation is vital to gain understanding of the dynamics of two solitons 
in integrable systems such as this one.  We find that the phase shift agrees with the one computed for the giant magnon \cite{Chen:2006gq} 
(up to non-logarithmic terms). This is perhaps surprising since both
the time delay and the energy for the single spike and the giant magnon are different.
Only after integrating the time delay with respect to the energy do both results agree as a function of the dressing parameters. 

 Another interesting conceptual advantage of the dressing method is that the ``basic'' or ``naked'' solution we need to dress to obtain the single
spike is a string wrapped around the equator of the sphere infinitely many
times. This confirms the idea of \cite{IK} that these solutions
can be thought of as excitations over that state. Identifying this state
in the field theory has however proved difficult.  Some ideas
in that respect including a
possible relation to the antiferromagnetic state of a spin chain are discussed in \cite{Zarembo:2005ur,Roiban:2006jt} and mentioned 
in \cite{IK}. However, in our mind the situation is not completely clear. It is quite interesting though, since we see that the solutions we 
consider have a very rich integrable dynamics.

 This paper is organized as follows:
 in section \ref{dressing} we review the main 
 ingredients of the dressing method in the $SU(2)$ case
relevant for our purposes.
In sections \ref{S2} and \ref{S3} we discuss the scattering solutions for strings moving on two and three-spheres 
respectively. In section \ref{phase shift} we compute the phase shift and compare to the case of the giant magnon. Finally, we
give our conclusions in section \ref{conclusions}.
 
\section{The dressing method}
\label{dressing}
 In this section we describe the basic idea of the dressing method with the main intention of establishing the notation (which is the same 
as in \cite{SV}) for the rest of the paper. Parameterizing $S^3$ with two complex coordinates $Z_{1,2}$ such that $|Z_1|^2+|Z_2|^2=1$, 
the Polyakov action for a string
moving on ${\mathbb{R}}_t\times S^3$ is
\beq
S =\frac{1}{2} \int d\sigma d\tau \left[ -(\partial_\tau t)^2+( \partial_\sigma t)^2 
       + \partial_\tau \bar{Z}_a \partial_\tau Z_a - \partial_\sigma \bar{Z}_a \partial_\sigma Z_a  
       -\Lambda\left(\bar{Z}_aZ_a-1\right)\right],
\eeq
where $a=1,2$. The variable $\Lambda$ is a Lagrange multiplier which enforces $|Z_1|^2+|Z_2|^2=1$.
This is supplemented by the conformal constraints, which, if we use the ansatz $t=\tau$ read
\beqa
 \partial_\tau \bar{Z}_a \partial_\tau Z_a + \partial_\sigma \bar{Z}_a \partial_\sigma Z_a &=& 1, \\
 \partial_\tau \bar{Z}_a \partial_\sigma Z_a + \partial_\sigma \bar{Z}_a \partial_\tau Z_a &=& 0.
\eeqa
 It turns out that if we define the $SU(2)$ matrix
\beq
g =\left(\begin{array}{cc}Z_1 &-iZ_2 \\-i\bar{Z}_2& \bar{Z}_1 \end{array}\right) ,
\eeq
then the
equations of motion and the constraints
can be written together in a compact form as
\beq
 \bar{\partial} \left(\partial g g^{-1}\right) + \partial\left(\bar{\partial}gg^{-1}\right) =0 ,
\label{geq}
\eeq
 where we defined $z=\half(\sigma-\tau)$, $\bar{z}=\half(\sigma+\tau)$ and $\partial=\partial_{z}$, $\bar{\partial}=\partial_{\bar{z}}$. 
Eq. (\ref{geq}) can be seen as the compatibility equation for the existence of a solution to the linear problem
\beq
 i \bar{\partial} \Psi = \frac{A\Psi}{1+\lambda}, \ \ \ \ \  i\partial\Psi=\frac{B\Psi}{1-\lambda},
\label{Psieq}
\eeq
where
\beq
A= i\bar{\partial}g g^{-1}, \ \ \ \ \ B = i\partial g g^{-1} ,
\eeq
and $\Psi$ is a two by two matrix.
 Given $g$ satisfying eq.(\ref{geq}) we can find $\Psi$ satisfying eq.(\ref{Psieq}) with the initial condition $\Psi(\lambda=0)=g$. Conversely,
if we have a solution $\Psi(\lambda,z,\bar{z})$ for given matrices $A(z,\bar{z})$ and $B(z,\bar{z})$ then $\Psi(0)$ is guaranteed
to satisfy eq.(\ref{geq}). Notice that
for this we need $A$ and $B$ independent of $\lambda$.

The basic point of the dressing
method is that given a solution $g$, from which $A$, $B$ and $\Psi$
can be determined,
one can then find a new solution by multiplying $\Psi$ by an appropriate matrix $\chi(\lambda)$: $\Psi\rightarrow \chi\Psi$.
Only for specific choices of $\chi$ will the product $\chi \Psi$ continue
to satisfy the desired eq.(\ref{Psieq}).
 In the examples we consider, the matrix $\chi$ takes the form
\beq
 \chi(\lambda) = 1 + \frac{\lambda_1-\bar{\lambda}_1}{\lambda-\lambda_1} P ,
\label{chi}
\eeq
where $\lambda_1$ an arbitrary complex parameter and $P$ is a projector defined as 
\beq
P = \frac{\Psi(\bar{\lambda}_1)ee^\dagger\Psi^{-1}(\lambda_1)}{e^\dagger \Psi^{-1}(\lambda_1)\Psi(\bar{\lambda}_1)e} ,
\eeq
in terms of a vector $e$ that can be set to $e = (1,1)$ without
loss of generality.
The rationale behind the choice (\ref{chi}) can be found in the 
references \cite{SV,dm1,dm2,dm3} together with a more complete explanation of the method. For us it suffices to know that,
given a solution $g(z,\bar{z})$ this method then provides a one (complex) parameter family of new solutions $g_{\lambda_1}(z,\bar{z})$
labeled by the complex number $\lambda_1$.
Successive applications of the dressing method can be used to generate more
complicated solutions depending on additional parameters $\lambda_2,\lambda_3,\ldots$.
The examples in the following sections should clarify how to apply this method in practice.

\section{Single spikes on $S^2$}
\label{S2}

 In this section we consider the solutions for strings on $S^2$ discussed in \cite{IK,Mo} and
 demonstrate how to
 generalize them to include many spikes. 
As a first step we rewrite the solution from \cite{IK},
which lives inside an $S^2 \subset S^5$,
in terms of six real embedding coordinates
$Y^i$ on $S^5$ as
\begin{eqnarray}
\label{spikeembedding}
Y^1 &=& {1 \over 2} (Z^1 + \overline{Z}^1) =
\sqrt{1 - \cos^2 \theta_1 \sech^2 \xi \over
1 + \cot^2 \theta_1 \tanh^2 \xi}
\left[ \cos x + \sin x \cot \theta_1 \tanh \xi \right],
\cr
Y^2 &=& {1 \over 2 i} (Z^1 - \overline{Z}^1) =
\sqrt{1 - \cos^2 \theta_1 \sech^2 \xi \over
1 + \cot^2 \theta_1 \tanh^2 \xi} 
\left[ \sin x - \cos x \cot \theta_1 \tanh \xi \right],
\cr
Y^3 &=& Z^2 = \cos \theta_1 \sech \xi,
\cr
Y^4 &=& Y^5 = Y^6 = 0,
\end{eqnarray}
where
$\theta_1$ is a parameter of the solution and
\begin{equation}
\xi = t \sec \theta_1 + x \tan \theta_1.
\end{equation}
Let us call the solution (\ref{spikeembedding}) $\vec{Y}_1$.

The simplest way to build a scattering state of this spike
$\vec{Y}_1$ with another spike
$\vec{Y}_2$ (which is given by the same formula as above but
with a different parameter $\theta_2$) 
is to use the formula \cite{Mikhailov:2005zd}
\begin{equation}
\label{magicformula}
\vec{Y}_{1,2} = \vec{Y}_0 + (\vec{Y}_1 - \vec{Y}_2)
{\vec{Y}_0 \cdot \vec{Y}_2 - \vec{Y}_0 \cdot \vec{Y}_1 \over
1 - \vec{Y}_1 \cdot \vec{Y}_2},
\end{equation}
where $\vec{Y}_0$ is the ``vacuum'' or ``naked'' solution, which in this context
is given by
\begin{equation}
Y_0 = ( \cos x, \sin x, 0, 0 , 0, 0 ) ,
\end{equation}
and describes
the string at rest winding infinitely many times around the equator
($\theta_0 = \pi/2$).

One can check directly that (\ref{magicformula}) satisfies the equations of motion
and Virasoro constraints
\begin{eqnarray}
- \partial_t^2 \vec{Y} + \partial_x^2 \vec{Y} + (- \partial_t \vec{Y}
\cdot \partial_t \vec{Y} + \partial_x \vec{Y} \cdot \partial_x \vec{Y})
\vec{Y} &=& 0,\cr
- \partial_t \vec{Y} \cdot \partial_x \vec{Y} &=& 0,\cr
\partial_t \vec{Y}
\cdot \partial_t \vec{Y} + \partial_x \vec{Y} \cdot \partial_x \vec{Y} &=&1,
\end{eqnarray}
and of course the embedding constraint $\vec{Y} \cdot \vec{Y} = 1$. Further application of the method
would allow us to construct solutions with more spikes but for our present purpose these two-spike solutions
are sufficient. 

It is important to note that one must first switch to the $(t,x)$ coordinates
\begin{equation}
\label{tausigma}
\tau = t \sec \theta_1, \qquad \sigma = x \sec \theta_1
\end{equation}
before applying the relation (\ref{magicformula}).  That is, 
from eq.(\ref{tausigma}) it is clear
that the $\sigma$ and $\tau$ coordinates of the solutions
$\vec{Y}_1$ and $\vec{Y}_2$ are normalized differently; they
cannot be combined using eq.(\ref{magicformula}) until we first switch to the common
$(t,x)$ coordinates. We now turn to the case of single spikes on $S^3$ where
the
solution has an additional angular momentum parameter.

\section{Single spikes on $S^3$ and scattering solutions via dressing}
\label{S3}

 As noted in the previous section, to
superpose spikes it is important to rescale $\sigma$ and $\tau$
so that the relation $t = \kappa \tau$ holds with $\kappa = 1$ for
all solutions.
Going to conformal gauge and using the same ideas as in the previous case we find
a solution depending on two parameters
$\theta_1$ and $\gamma_1$.
If we now use the parameterization
\beqa
 Z_1 &=& Y_1 + i Y_2 = \sin\theta e^{i \phi_1} , \\
  Z_2 &=& Y_3 + i Y_4 = \cos\theta e^{i \phi_2} ,\\
   Z_3 &=& Y_5 + i Y_6 =0 ,
\eeqa
of $S^3$ in terms of three angles $(\theta,\phi_1,\phi_2)$, the solution is
\beqa
 \cos\theta &=& \frac{\cos\theta_1}{\cosh\xi}  ,\\
 \phi_1 &=& \sigma-\arctan\left(\frac{\cos\theta_1}{\sin\theta_1}\tanh\xi\right) ,\\
 \phi_2 &=& \frac{\sin\gamma_1}{1-\cos^2\gamma_1\sin^2\theta_1} \left(\sigma+\tau\,\cos\gamma_1\sin\theta_1\right) ,\\
 \xi &=& \frac{\cos\theta_1\cos\gamma_1}{1-\cos^2\gamma_1\sin^2\theta_1}\left(\tau+\sigma\,\cos\gamma_1\sin\theta_1\right).
\eeqa
This solution has one more conserved angular momentum and lives in an
$S^3 \subset S^5$. Its properties were studied in \cite{IK}.
Now we show that this solution follows from the infinitely wrapped string by using the same dressing method that in \cite{SV} was used for giant magnons.

{}From now on we set $Z_3=0$ and consider, as mentioned, solutions on ${\mathbb{R}}_t\times S^3$ where the $S^3$ is parameterized by $Z_{1,2}$ 
with $|Z_1|^2+|Z_2|^2=1$. This allows us to apply the ideas described in section \ref{dressing} directly.   
 We start from the infinitely wrapped string solution\footnote{The giant magnon is similarly constructed from the $S^3$ solution $Z_1=e^{i\tau}$, $Z_2=0$.
The full ${\mathbb{R}}_t \times S^3$ solution for the spike is however not a simple $\sigma\leftrightarrow \tau$ interchange of the magnon since
$t=\tau$ for both the magnon and the spike. Equivalently we can say that we interchange $t=\tau$ for $t=\sigma$. It is easy to see that, in conformal 
gauge and for a metric ${\mathbb{R}}_t\times S^3$ this maps solutions into solutions. We thank A. Tseytlin for this last comment. }
\beq
 Z_1 = e^{i\sigma} , \ \ \ Z_2 = 0 .
\eeq 
The embedding into $SU(2)$ with $z=\frac{1}{2}(\sigma-t)$ and 
$\bz=\frac{1}{2}(\sigma+t)$ is given by
\beq
 g = \left( \begin{array}{cc} e^{i(z+\bz)} & 0 \\ 0 & e^{-i(z+\bz)} \end{array} \right) ,
\eeq
which leads to (using the notation from \cite{SV} or section \ref{dressing})
\beq
 A=B = \left( \begin{array}{cc} -1 & 0 \\ 0 & 1 \end{array} \right).
\eeq
The solution of the corresponding linear problem (\ref{Psieq}) is
\beq
 \Psi = \left( \begin{array}{cc} e^{iZ(\lambda)} & 0 \\ 0 & e^{-iZ(\lambda)} \end{array} \right), \ \ \ 
   Z(\lambda)= \frac{z}{1-\lambda}+\frac{\bz}{1+\lambda}.
\eeq
 Taking the constant vector $e=(1,1)$ we obtain the projection operator
\beq
P=\frac{1}{1+e^{2i(Z(\lambda_1)-Z(\bar{\lambda}_1))}} 
\left( \begin{array}{cc} 1 & e^{2iZ(\lambda_1)} \\ 
	e^{-2iZ(\bar{\lambda}_1)} & e^{2i(Z(\lambda_1)-Z(\bar{\lambda}_1))} \end{array} \right).
\eeq
The method then gives a family of new solutions (after including a
normalization factor to maintain $\det\Psi(\lambda=0) =1$)
\beq
g_{\lambda_1} =  \Psi_{\lambda_1}(0) = \sqrt{\frac{\lambda_1}{\bar{\lambda}_1}}\left[ 1-\frac{\lambda_1-\bar{\lambda}_1}{\lambda_1} P \right] 
                                           \Psi(\lambda=0).
\eeq
{}We can now read off the coordinates $Z_{1,2}$ of this solution, finding
\beqa
Z_1 &=& \frac{e^{i\sigma}}{\sqrt{\lambda_1\bar{\lambda}_1}} \frac{\lambda_1 e^{-2iZ(\bar{\lambda}_1)}+\bar{\lambda}_1 
            e^{-2iZ(\lambda_1)}}{e^{-2iZ(\lambda_1)}+e^{-2iZ(\bar{\lambda}_1)}}, \label{Z_1} \\
Z_2 &=& \frac{e^{-i\sigma}}{\sqrt{\lambda_1\bar{\lambda}_1}} \frac{i(\bar{\lambda}_1- \lambda_1)}{e^{-2iZ(\lambda_1)}
            +e^{-2iZ(\bar{\lambda}_1)}} \label{Z_2}.
\eeqa
The energy and angular momentum can be computed as
\beqa
\varepsilon = E-T\Delta \phi 
  &=& \frac{\sqrt{\lambda}}{2\pi} \int_{-\infty}^{+\infty}d\sigma\left(1 - \partial_\sigma {\rm Im}\left[\log Z_1 \right]\right) ,\label{EmiD} \\
J_i &=& \frac{\sqrt{\lambda}}{2\pi} \int_{-\infty}^{+\infty}d\sigma\, {\rm Im}\left[\bZ_i \partial_t Z_i \right], \quad i = 1, 2 \label{angmomJ} 
\eeqa
where $\l$ is the 't Hooft coupling. The energy itself
$E=\frac{\sqrt{\lambda}}{2\pi} \int_{-\infty}^{\infty} d\sigma$ is infinite but
the excitation energy $\varepsilon$ above the infinitely wrapped
string ``vacuum'' 
is finite\footnote{This has some similarity with the situation discussed 
in \cite{wrapped}.}.   Henceforth we will usually refer
to $\varepsilon$ as the energy of the solution.

Substituting (\ref{Z_1}) into (\ref{EmiD}) and choosing
to parameterize $\lambda_1$ via
\beq
 \lambda_1 = r e^{ip/2},
\eeq
where $0<r<\infty$, $-\pi < \frac{p}{2} < \pi$,  we obtain
\begin{equation}
\varepsilon = \frac{\sqrt{\lambda}}{\pi} \left[ \frac{\pi}{2} - \left|\left|\frac{p}{2}\right|-\frac{\pi}{2}\right| \right] =\left\{
      \begin{array}{ll}
        \frac{\sqrt{\lambda}}{\pi}\left|\frac{p}{2}\right|,
	& \quad \textrm{if}\ \left|\frac{p}{2}\right| < \frac{\pi}{2} \\ &\\
        \frac{\sqrt{\lambda}}{\pi}\left(\pi-\left|\frac{p}{2}\right| \right), 
	& \quad \textrm{if}\ \left|\frac{p}{2}\right| > \frac{\pi}{2} \\
      \end{array} \right.
\label{EmiD2}
\end{equation}
which is plotted in fig.\ref{evsp} for convenience. Notice that the energy is always positive. 
\FIGURE{\epsfig{file=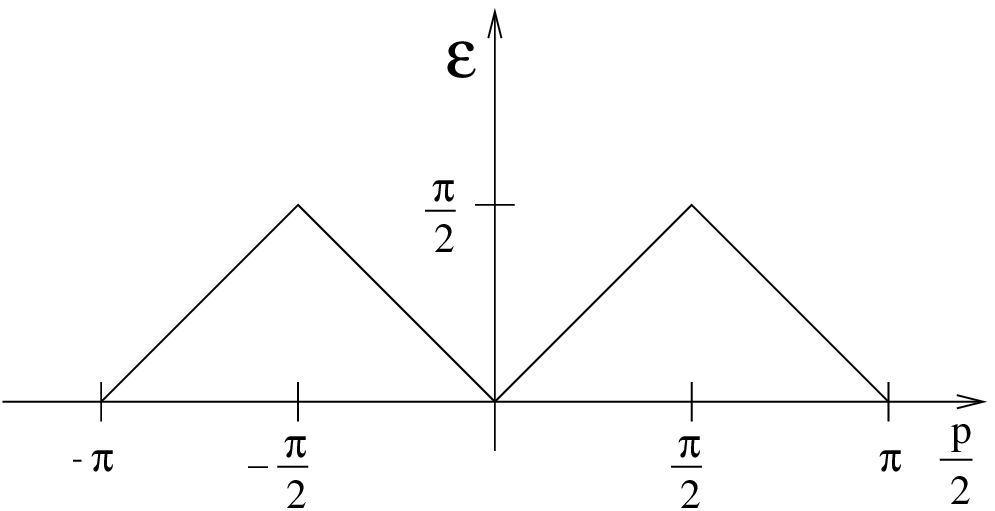, height=5cm}
\caption{Excitation energy
of the single spike solution
as a function of the parameter $\frac{p}{2}$.}
\label{evsp}
}

Similarly, equations (\ref{angmomJ}), (\ref{Z_1}), and (\ref{Z_2}) lead to
\beqa
J_1 &=& \sg\left(\sin p\right)\, \frac{\sqrt{\l}}{i4\pi}
\left[ \left( \l_1-\frac{1}{\l_1} \right) -\left( \bar{\l}_1-\frac{1}{\bar{\l}_1} \right) \right] \nonumber \\ 
 &=& \sg\left(\sin p\right)\, \frac{\sqrt{\l}}{\pi} \frac{1+r^2}{2r}\sin \frac{p}{2}, \label{angJ1} \\ 
J_2 &=& \sg\left(\sin p\right)\, \frac{\sqrt{\l}}{i4\pi}
\left[ \left( \l_1+\frac{1}{\l_1} \right) -\left( \bar{\l}_1+\frac{1}{\bar{\l}_1} \right) \right] \nonumber \\ 
&=& \sg\left(\sin p\right)\, \frac{\sqrt{\l}}{\pi} \frac{r^2-1}{2r}\sin \frac{p}{2}. \label{angJ2}  
\eeqa
Upon
eliminating $r$ in the above
equations we find that the two angular momenta are related by
\beq
 J_1 = \sqrt{J_2^2 + \frac{\l}{\pi^2} \sin^2 \frac{p}{2}},
 \label{JJ}
\eeq
One can check that equations (\ref{EmiD2}) and (\ref{JJ}) agree with the expressions obtained
in \cite{IK} for the single spike solution when we identify
the parameter $\bar{\theta}$ there with
$p/2=\bar{\theta}$. In \cite{IK}
only
the case $0< \bar{\theta} <\frac{\pi}{2}$ was considered, whereas here
we can take $-\pi<\bar{\theta}<\pi$. 
Extending the range of $\bar{\theta}$ includes solutions which are related by reflections to the solutions in 
the $0< \bar{\theta} <\frac{\pi}{2}$ range and therefore are not truly
independent (they are just spikes moving in the opposite direction).
Shortly we will superpose solutions and it will be important
to consider single spikes in the full range $-\pi<\bar{\theta}<\pi$.

 As a reminder, in terms of the same dressing parameter $\lambda_1=r e^{i\frac{p}{2}}$, for the giant magnon solution of \cite{HM}, 
the energy and angular momentum are given by (see \cite{SV}),
\begin{eqnarray}
E^{(\rm{mag})} &=& \frac{\sqrt{\l}}{i4\pi} \sg(\sin\frac{p}{2}) 
\left[ \left( \l_1-\frac{1}{\l_1} \right) -\left( \bar{\l}_1-\frac{1}{\bar{\l}_1} \right) \right] 
 = \frac{\sqrt{\l}}{\pi} \frac{1+r^2}{2r}\left|\sin \frac{p}{2}\right|, \label{Emag} \\ 
J_2^{(\rm{mag})} &=& -\frac{\sqrt{\l}}{i4\pi} \sg(\sin\frac{p}{2})
\left[ \left( \l_1+\frac{1}{\l_1} \right) -\left( \bar{\l}_1+\frac{1}{\bar{\l}_1} \right) \right] 
= \frac{\sqrt{\l}}{\pi} \frac{1-r^2}{2r}\left|\sin \frac{p}{2}\right|. \label{angJ2mag}  
\end{eqnarray}
Eliminating $r$ in these equations leads to the relation
\beq
E^{(\rm{mag})} = \sqrt{\{J_2^{(\rm{mag})}\}^2 + \frac{\l}{\pi^2} \sin^2 \frac{p}{2}}.
 \label{JJmag}
\eeq

Going back to the spike solution and repeating the dressing method again, from a single two-charge soliton we obtain
a two spike solution
\beqa
Z_1 &=& \frac{e^{i \sigma}}{2 |\lambda_1 \lambda_2|} 
\frac{R + |\lambda_1|^2 \lambda_{1 \bar{1}} \lambda_{2 \bar{2}} e^{+i (v_1 - v_2)}
+ |\lambda_2|^2 \lambda_{1 \bar{1}} \lambda_{2 \bar{2}} e^{-i (v_1 - v_2)}}
{\lambda_{12} \lambda_{\bar{1} \bar{2}} \cosh(u_1 + u_2)
+ \lambda_{1 \bar{2}} \lambda_{\bar{1} 2} \cosh(u_1 - u_2)
+ \lambda_{1 \bar{1}} \lambda_{2 \bar{2}} \cos(v_1 - v_2)}, \nonumber \\ \\
Z_2 &=& \frac{-i}{2 |\lambda_1 \lambda_2|}
\frac{ \lambda_{1 \bar{1}} e^{i v_1} \left[
\lambda_{12} \lambda_{\bar{1} 2}  \bar{\lambda}_2 e^{+u_2}
+ \lambda_{\bar{1} \bar{2}} \lambda_{1 \bar{2}} \lambda_2 e^{-u_2}
\right] + (1 \leftrightarrow 2)}
{\lambda_{12} \lambda_{\bar{1} \bar{2}} \cosh(u_1 + u_2)
+ \lambda_{1 \bar{2}} \lambda_{\bar{1} 2} \cosh(u_1 - u_2)
+ \lambda_{1 \bar{1}} \lambda_{2 \bar{2}} \cos(v_1 - v_2)},  \nonumber \\  
\label{ZZ_2}
\eeqa
where
\beq
  R = \lambda_{12} \lambda_{\bar{1} \bar{2}} \left[
  \lambda_1 \lambda_2 e^{+u_1 + u_2} + \bar{\lambda}_1 \bar{\lambda}_2 e^{-u_1 - u_2}
  \right]
  + \lambda_{\bar{1} 2} \lambda_{1 \bar{2}} \left[
  \lambda_1 \bar{\lambda}_2 e^{+u_1 - u_2} + \bar{\lambda}_1 \lambda_2 e^{-u_1 + u_2}
  \right], 
\eeq
and
\beqa
  u_i &=& i (Z(\lambda_i) - Z(\bar{\lambda}_i)), \nonumber \\
  v_i &=& Z(\lambda_i) + Z(\bar{\lambda}_i) - \sigma, \quad i=1,2.
  \label{uv}
\eeqa
The following shorthand notation has been used:
\beq
\l_{12}=\l_1-\l_2,  \quad \l_{1\bar{2}}=\l_1-\bar{\l}_2, \quad {\rm etc.}
\eeq
Substituting $\l_i=r_i e^{ip_i/2}$, we can express $|Z_2|$ as a function of $t$ and $\sigma$. To gain some understanding of the solution we plot
$|Z_2|$ as a function of $\sigma$ for different values of $t=\tau$ in fig.\ref{scattering}. At an early time the spikes are far apart. However 
they come close, eventually scattering and separating from each other. At late times, the profile again describes two separated solitons, 
the only evidence of the scattering being that their positions are shifted with respect to what they would be if they had moved past each other
with constant velocity. 
 Numerically we can compute the shift and from there the time delay, namely the difference between the time at which the soliton arrives to a 
given point and the time at which it would have arrived if it had not met the other soliton in the way. This serves to illustrate the analytical 
calculations we perform in the next section and the numerical results also
provide a useful check.
 As a final point, since when $t\rightarrow -\infty$ the solitons are far apart, the energy and angular momenta of the solution is simply the
sum of the ones for each soliton separately. 

\FIGURE{\epsfig{file=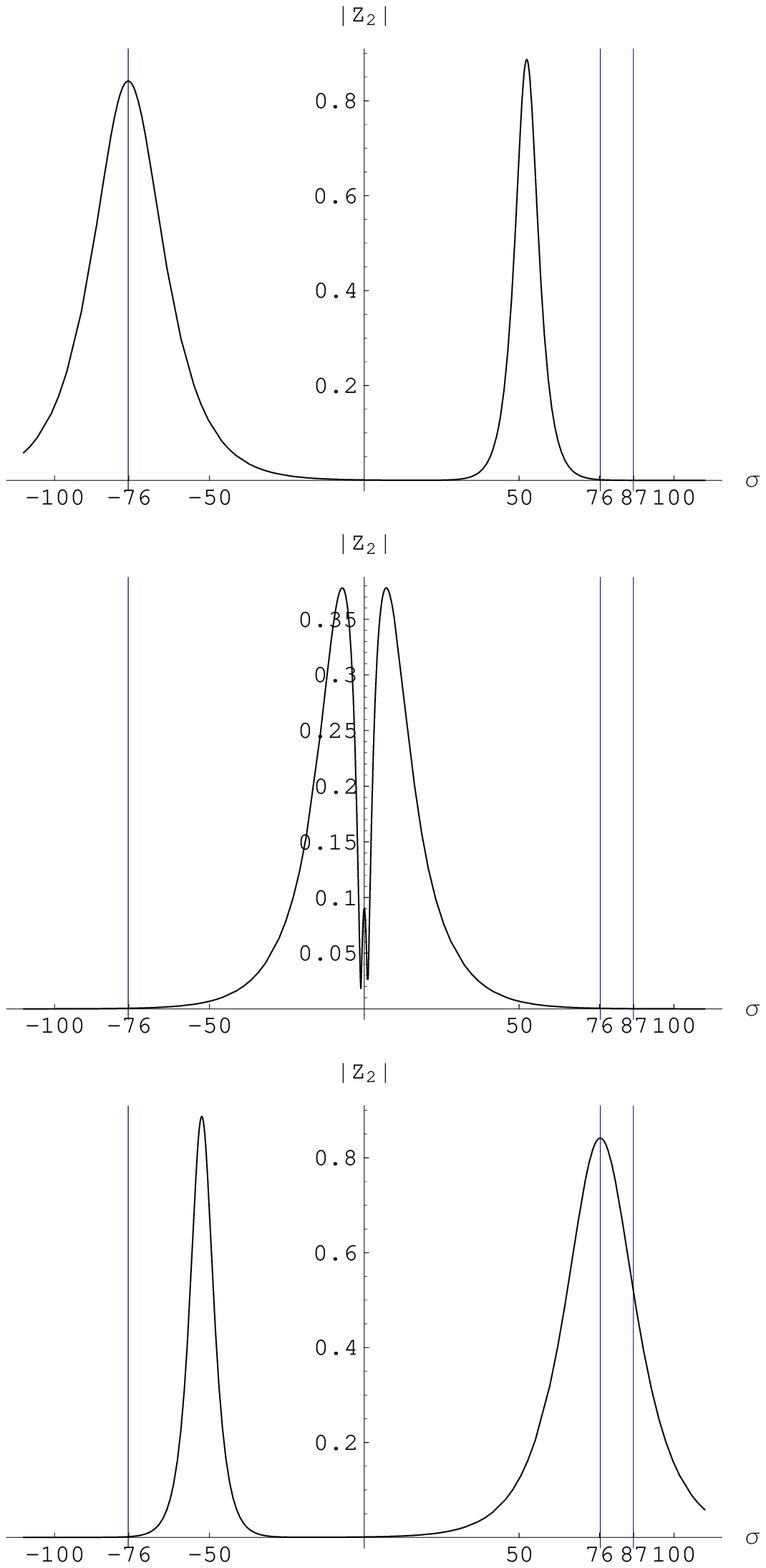, width=9.5cm}
\caption{$|Z_2|$ as a function of $\sigma$ with
the choice of parameters
$r_1=0.24,\ p_1=2.0,\ r_2=0.50,$
and $p_2=4.1$ at $t=-20,\ 0,\ 20$, from top to bottom, respectively.
The line at $\sigma=87$ is the expected final location of the first soliton 
had it not encountered the second soliton.
The velocity and distance shift of the first soliton are, $V_1=4.08$ and $\delta \sigma_1=10.7$, respectively, which agree with the analytic
 result given in the text.}
\label{scattering}
}

\section{Scattering phase shift}
\label{phase shift}

To compute the time delay, we first consider a single soliton as given by eq.(\ref{Z_2}). Using this together with eq.(\ref{uv}), we obtain,
\begin{equation}
	\left|Z_2\right| \propto \frac{1}{\cosh u_1} .
\end{equation}
 This shows that the position of the soliton, namely, the maximum of $|Z_2|$, is determined by equating $u_1=0$. In particular this implies
that the soliton moves with constant velocity
\beq
V_i = \frac{1+\l_i\bar{\l}_i}{\l_i+\bar{\l}_i}= \frac{1+r_i^2}{2 r_i\cos\frac{p_i}{2}}, \quad i=1,2.
\eeq
We can make this explicit by writing
\begin{equation}
	u_i=\frac{i(\l_i^2-\bl_i^2)}{\left|1-\l_i^2\right|^2}(\sigma-V_i t),
	\quad i=1,2.
	\label{u_sigma}
\end{equation}
 Now we compute the time delay that particle 1 experiences as it goes through the other particle assuming particle 1 starts from the left.
They can be moving toward each other ($V_2<0<V_1$) or in the same direction ($0<V_2<V_1$).  
 To find the distance shift that particle 1 experiences, we first compute the location of the particle when $\sigma, t \rightarrow -\infty$. 
We can obtain the location by computing the extremum of eq.(\ref{ZZ_2}) with respect 
to $u_1$ and then equate the result with eq.(\ref{u_sigma}) in this limit. 
The initial location of particle one at $t=-t_0$ is,
\begin{equation}
	\sigma_{i}(t=-t_0)=\sigma_1(\sigma,t\rightarrow -\infty)=-V_1 t_0+\sg\left(\sin p_2\right)\, 
	\frac{i\left|1-\bl_1^2\right|^2}{2(\l_1^2-\bl_1^2)}
\log\left[\frac{\l_{12}\l_{\bar{1}\bar{2}}}{\l_{1\bar{2}}\l_{\bar{1}2}}\right].
	\label{sigma_neg}
\end{equation}
Similarly at the other limit, we obtain the final location of the particle at
$t=+t_0$,
\begin{equation}
\sigma_{f}(t=+t_0)=\sigma_1(\sigma,t\rightarrow +\infty)=V_1t_0-\sg\left(\sin p_2\right)\, 
\frac{i\left|1-\bl_1^2\right|^2}{2(\l_1^2-\bl_1^2)}
\log\left[\frac{\l_{12}\l_{\bar{1}\bar{2}}}{\l_{1\bar{2}}\l_{\bar{1}2}}\right],
\label{sigma_pos}
\end{equation}
whereas the expected location of particle 1 in the limit, from eq.(\ref{sigma_neg}), is
\begin{equation}
\sigma_{exp}=\sigma_{i}(t=-t_0)+2V_1t_0=V_1t_0+\sg\left(\sin p_2\right)\, 
	        \frac{i\left|1-\bl_1^2\right|^2}{2(\l_1^2-\bl_1^2)}
		\log\left[\frac{\l_{12}\l_{\bar{1}\bar{2}}}{\l_{1\bar{2}}\l_{\bar{1}2}}\right] .
\end{equation}
Thus, the distance shift that particle 1 experiences as it goes through the
other particle starting from the left is 
\beqa
\delta \sigma_1 &=& \sigma_{exp}-\sigma_{f}\\
&=& \sg\left(\sin p_2\right)\, \frac{i(1-\l_1^2)(1-\bar{\l}_1^2)}{(\l_1^2-\bar{\l}_1^2)} 
\log\left[\frac{\l_{12}\l_{\bar{1}\bar{2}}}{\l_{1\bar{2}}\l_{\bar{1}2}} \right] \\
&=& \sg\left(\sin p_2\right)\, \frac{1+r_1^4-2r_1^2\cos p_1}{2 r_1^2 \sin p_1} 
\log\left[\frac{r_1^2+r_2^2-2r_1 r_2 \cos\frac{p_1-p_2}{2}}{r_1^2+r_2^2-2r_1 r_2\cos \frac{p_1+p_2}{2}} \right].
\eeqa
Finally, the time delay for particle 1 becomes
\begin{eqnarray}
	\Delta T_1&=& \frac{\delta \sigma_1}{V_1}
	=  \sg\left(\sin p_2\right)\, \frac{i(1-\l_1^2)(1-\bar{\l}_1^2)}{(\l_1-\bar{\l}_1)(1+\l_1\bar{\l}_1)} 
  \log\left[\frac{\l_{12}\l_{\bar{1}\bar{2}}}{\l_{1\bar{2}}\l_{\bar{1}2}} \right] \nonumber \\
	&=& \sg\left(\sin p_2\right)\, \frac{1+r_1^4-2r_1^2\cos p_1}{2 r_1(1+r_1^2) \sin \frac{p_1}{2}} 
\log\left[\frac{r_1^2+r_2^2-2r_1 r_2 \cos\frac{p_1-p_2}{2}}{r_1^2+r_2^2-2r_1 r_2\cos \frac{p_1+p_2}{2}} \right].
\label{dT}
\end{eqnarray}

For comparison, the velocity, position shift and time delay for the scattering of giant magnons are
\begin{eqnarray}
V_i^{(\rm{mag})} &=& \frac{\l_i+\bar{\l}_i}{1+\l_i\bar{\l}_i}=\frac{2 r_i\cos\frac{p_i}{2}}{1+r_i^2}, \\
\delta \sigma_1^{(\rm{mag})}
&=& \sg(\sin\frac{p_2}{2}) i\frac{(1-\l_1^2)(1-\bar{\l}_1^2)}{(\l_1-\bar{\l}_1)(1+\l_1\bar{\l}_1)} 
  \log\left[\frac{\l_{12}\l_{\bar{1}\bar{2}}}{\l_{1\bar{2}}\l_{\bar{1}2}} \right] \nonumber \\
&=& \sg(\sin\frac{p_2}{2}) \frac{1+r_1^4-2r_1^2\cos p_1}{2 r_1(1+r_1^2) \sin \frac{p_1}{2}} 
\log\left[\frac{r_1^2+r_2^2-2r_1 r_2 \cos\frac{p_1-p_2}{2}}{r_1^2+r_2^2-2r_1 r_2\cos \frac{p_1+p_2}{2}} \right], \\
\Delta T_1^{(\rm{mag})} 
&=&  \sg(\sin\frac{p_2}{2}) i\frac{(1-\l_1^2)(1-\bar{\l}_1^2)}{(\l_1^2-\bar{\l}_1^2)} 
\log\left[\frac{\l_{12}\l_{\bar{1}\bar{2}}}{\l_{1\bar{2}}\l_{\bar{1}2}} \right] \\
&=& \sg(\sin\frac{p_2}{2}) \frac{1+r_1^4-2r_1^2\cos p_1}{2 r_1^2 \sin p_1} 
\log\left[\frac{r_1^2+r_2^2-2r_1 r_2 \cos\frac{p_1-p_2}{2}}{r_1^2+r_2^2-2r_1 r_2\cos \frac{p_1+p_2}{2}} \right].
\end{eqnarray}
Note that the distance shift and time delay are interchanged compared
to those of the spikes.

We can now compute the phase shift with the formula
\begin{equation}
	\left(\frac{\partial \delta_1}{\partial \varepsilon_1}\right)_{J_2}
	= \Delta T_1 ,
\end{equation}
where the angular momentum, $J_2$ must be fixed when the above equation is integrated. It is easy to check that the integral is
given by the same function $\Theta$ that appears in the giant magnon calculation of \cite{Chen:2006gq}
\begin{equation}
	\Theta \left(\l_1,\bar{\l}_1,\l_2,\bar{\l}_2\right) = \frac{\sqrt{\l}}{2\pi}
	\left[ K\left(\l_1,\l_2\right)+K\left(\bar{\l}_1,\bar{\l}_2\right)-K\left(\l_1,\bar{\l}_2\right)-K\left(\bar{\l}_1,\l_2\right) \right] ,
	\label{dorey}
\end{equation}
where the function $K$ is given by
\begin{equation}
	K\left(X,Y\right)= \left[ \left( X+\frac{1}{X} \right)-\left( Y+\frac{1}{Y} \right) \right] \log(X-Y) .
\end{equation}
Indeed, using eqns.(\ref{EmiD2}), (\ref{angJ1}) and (\ref{angJ2}) we can compute
\beq
\left( \frac{\partial \l_1}{\partial \varepsilon_1} \right)_{J_2} = 
 \sg\left(\sin p_1\right)\, \frac{2\pi}{\sqrt{\l}} \frac{i\l_1^2(1-\bl_1^2)}{(\l_1-\bl_1)(1+\l_1\bl_1)}, \ \ \ \ \ 
\left( \frac{\partial \bl_1}{\partial \varepsilon_1} \right)_{J_2}=\overline{ \left( \frac{\partial \l_1}{\partial \varepsilon_1} \right)}_{J_2} ,
\eeq
and then use them to differentiate $\Theta$ in eq.(\ref{dorey}) with respect to $\varepsilon_1$. We obtain,
\beqa
    \left( \frac{\partial \Theta}{\partial \varepsilon_1}\right)_{J_2}
    &=& -\sg(\sin p_1\sin p_2) \Delta T_1+ \sg\left(\sin p_1\right)\, \frac{i(\l_2-\bar{\l}_2)}{\l_2\bar{\l}_2} \\
    &=&   -\sg(\sin p_1\sin p_2) \frac{\partial \delta_1}{\partial \varepsilon_1}
          -\sg(\sin p_1\sin p_2) \frac{2\pi}{\sqrt{\l}} \left( J_1^{(2)}-J_2^{(2)} \right),
\eeqa
where $J_1^{(2)}$ and $J_2^{(2)}$ are the angular momenta of the second soliton.

The phase shift that particle 1 experiences as it goes through particle 2 is then, 
\begin{eqnarray}
\delta_1
&=& -\sg(\sin p_1\sin p_2) \Theta\left(\l_1,\bar{\l}_1,\l_2,\bar{\l}_2\right) 
               + \sg\left(\sin p_2\right)\, \frac{i(\l_2-\bar{\l}_2)}{\l_2\bar{\l}_2} \varepsilon_1 \\
&=& -\sg(\sin p_1\sin p_2) \Theta \left(\l_1,\bar{\l}_1,\l_2,\bar{\l}_2\right) 
               - \frac{2\pi}{\sqrt{\l}} \left( J_1^{(2)}-J_2^{(2)} \right) \varepsilon_1.
	\label{p_delay_spik}
\end{eqnarray}
The phase shift for the giant magnon is known from \cite{Chen:2006gq} to be,  
\begin{equation}
	\delta_1^{(\rm{mag})}=
	-\sg(\sin \frac{p_1}{2}\sin \frac{p_2}{2}) \Theta \left(\l_1,\bar{\l}_1,\l_2,\bar{\l}_2\right) 
               - \left( \varepsilon_2+J_2^{(\rm{mag} ,2)} \right) p_1 .
	\label{p_delay_mag}
\end{equation}
 Remarkably we see a perfect parallel between the two results,
 despite the fact that intermediate steps are different. Ignoring a possible sign, the phase shift is the same up to non-logarithmic terms.
 The non-logarithmic terms are in any case non-universal.
For the scattering of giant magnons \cite{HM} such terms were absorbed by redefining the coordinate $\sigma$ in agreement with the expectations 
from the spin chain side. Here we do not clearly know the spin chain description of the system so we do not have any guide about how to treat the 
non-logarithmic terms. We leave this point for future understanding when the dual spin chain system is better known. 
 A clue in this respect is that the giant magnon phase appears as the strong-coupling limit of the scattering phase proposed in \cite{AFS} from 
field theory considerations. Presumably, since the phase for scattering of single spikes is the same as for giant magnons, the AFS phase also
plays a role in understanding these new solutions.

\section{Conclusions}
\label{conclusions}

 In this paper we have shown that the recently studied single spike solutions follow very simply as excitations of the string wrapped around the equator
by applying the dressing method. This allows us to find more generic solutions where the profile of the string is not rigid. In particular, we found
a solution
describing the scattering of two spikes and calculated the corresponding
phase shift. Perhaps surprisingly the result is the same
as for the giant magnon when written in terms of the dressing parameters,
even though intermediate steps in the calculation were quite different. This shows that the same integrable structure lies behind both and should perhaps
give a clue to the spin chain description of the single spike solutions which is still missing. 

\section*{Acknowledgments}

 We are grateful to M. Abbott, I. Aniceto, H.Y. Chen, A. Jevicki, C. Kalousios
 and A. Tseytlin for comments and discussions. The work of R.I. was supported in part by the Purdue Research Foundation 
and that of M.K. in part by NSF under grant PHY-0653357. 
The research of
MS is supported by NSF grant PHY-0610259 and by an OJI award
under DOE grant DE-FG02-91ER40688.
The research of AV is supported by NSF CAREER Award PHY-0643150 and
by DOE grant DE-FG02-91ER40688.

\section*{Note Added}

 While this paper was being written, \cite{note} appeared which has some overlap with section \ref{S3}. Also, in that paper an interesting 
field theory interpretation of these solutions is proposed.
The phase shift we compute here might be useful to test that proposal.

\end{document}